\def\frs{\raise.5ex\hbox{$\scriptstyle\numer$}\!/\!_\denom}\def
\frss{\raise.2ex\hbox{$\scriptscriptstyle\numer$}\!/\!_\denom
}\def\frac#1#2{{\def\numer{#1}\def\denom{#2}\mathchoice{\frs
}{\frs}{\frss}{\frss}}}\def\onehalf
{\frac12}\def\registered
{{\ooalign{\hfil\raise.05ex\hbox{\sevenrm R}\hfil\crcr\mathhexbox
20D}}}\def\missing#1{}\tolerance=3000\def\twtabskip{4pt
plus 72pt}\def\secskip{.1\vsize}\newskip\footmargin\def\textindent
#1{\ifdim\leftskip=0pt\leftskip=\footmargin\fi\indent\llap{\hbox
to\leftskip{#1\hss}}\ignorespaces}\def\tableft
{\advance\dimen1by\dimen0\dimen0=\wd0\hbox to-\dimen1{\hss}\unhbox
0}\def\tabcentre{\advance\dimen1by\dimen0\dimen0=\wd0\divide
\dimen0by2\advance\dimen1by\dimen0\hbox to-\dimen1{\hss}\unhbox
0}\def\tabright{\advance\dimen1by\dimen0\dimen0=\wd0\advance
\dimen1by\dimen0\dimen0=0pt\hbox to-\dimen1{\hss}\unhbox0}\def
\tabnone{\dimen0=\wd0\unhbox0}\def\textsuper
#1{{\setbox0=\hbox{#1\vphantom I}\raise.5\ht0\box0}}\def\textsub
#1{{\setbox0=\hbox{#1\vphantom I}\lower.5\ht0\box0}}\newdimen
\linedif\def\advprev{\advance\linedif by\prevdepth\prevdepth
=\linedif}\def\baselines#1{\linedif=#1\advance\linedif by-\ht
\strutbox\setbox\strutbox=\hbox{\vrule height\ht\strutbox depth\linedif
width0pt}\ifvmode\advance\baselineskip by-#1\linedif=-\baselineskip
\ifdim0pt<\linedif\advprev\else\vskip-\linedif\fi\else\strut
\fi\baselineskip=#1}\def\normalspacing{\baselines\normalbaselineskip
}\def\mathspacing{\normalspacing\ifvmode\linedif=\normalbaselineskip
\advprev\fi}\newdimen\headerdepth\def\makeheadline{\vbox to
0pt{\vskip-\headerdepth\vskip-\parskip\the\headline\vss}\nointerlineskip
}\def\makefootline{\vbox{\the\footline}}\newfam\bifam
\font\fontaaa=cmr10 at 12pt
\font\fontaab=cmti10 at 12pt
\font\fontaac=cmtt10 at 12pt
\font\fontaad=cmbx10 at 12pt
\font\fontaae=cmti10 at 12pt
\font\fontaaf=cmr10 at 18pt
\font\fontaag=cmtt10 at 18pt
\font\fontaah=cmbx10 at 18pt
\font\fontaai=cmti10 at 18pt
\let\fontaaj=\tenrm
\font\fontaba=cmr5 at 6.625pt
\let\fontabb=\fiverm
\let\fontabc=\tenit
\font\fontabd=cmti10 at 6.625pt
\font\fontabe=cmti10 at 5pt
\let\fontabf=\teni
\font\fontabg=cmmi10 at 6.625pt
\font\fontabh=cmmi10 at 5pt
\let\fontabi=\tensy
\font\fontabj=cmsy10 at 6.625pt
\font\fontaca=cmsy10 at 5pt
\let\fontacb=\tentt
\let\fontacc=\tenbf
\font\fontacd=cmbx5 at 6.625pt
\let\fontace=\fivebf
\let\fontacf=\tenit
\font\fontacg=cmti10 at 6.625pt
\font\fontach=cmti10 at 5pt
\font\fontaci=cmr10 at 20pt
\font\fontacj=cmtt10 at 20pt
\font\fontada=cmbx10 at 20pt
\font\fontadb=cmti10 at 20pt
\def\fontsetaa{\let\rm=\fontaaa\let\it=\fontaab\let\tt=\fontaac
\let\bf=\fontaad\let\bi=\fontaae\setbox\strutbox=\hbox{\vrule
height10.05pt depth3.25pt width0pt}\normalbaselineskip=13.3pt
\normallineskip=1.3pt\normallineskiplimit=1.3pt\normalspacing
\rm}
\def\fontsetab{\let\rm=\fontaaf\let\tt=\fontaag\let\bf=\fontaah
\let\bi=\fontaai\setbox\strutbox=\hbox{\vrule height15.05pt
depth4.85pt width0pt}\normalbaselineskip=19.9pt\normallineskip
=1.9pt\normallineskiplimit=1.9pt\normalspacing\rm}
\def\fontsetac{\def\rm{\fam0\fontaaj}\textfont0=\fontaaj\scriptfont
0=\fontaba\scriptscriptfont0=\fontabb\def\it{\fam\itfam\fontabc
}\textfont\itfam=\fontabc\scriptfont\itfam=\fontabd\scriptscriptfont
\itfam=\fontabe\textfont1=\fontabf\scriptfont1=\fontabg\scriptscriptfont
1=\fontabh\textfont2=\fontabi\scriptfont2=\fontabj\scriptscriptfont
2=\fontaca\def\tt{\fam\ttfam\fontacb}\textfont\ttfam=\fontacb
\def\bf{\fam\bffam\fontacc}\textfont\bffam=\fontacc\scriptfont
\bffam=\fontacd\scriptscriptfont\bffam=\fontace\def\bi{\fam
\bifam\fontacf}\textfont\bifam=\fontacf\scriptfont\bifam=\fontacg
\scriptscriptfont\bifam=\fontach\setbox\strutbox=\hbox{\vrule
height8.35pt depth2.75pt width0pt}\normalbaselineskip=11.1pt
\normallineskip=1.1pt\normallineskiplimit=1.1pt\normalspacing
\rm}
\def\fontsetad{\let\rm=\fontaci\let\tt=\fontacj\let\bf=\fontada
\let\bi=\fontadb\setbox\strutbox=\hbox{\vrule height16.7pt depth
5.5pt width0pt}\normalbaselineskip=22.2pt\normallineskip=2.2pt
\normallineskiplimit=2.2pt\normalspacing\rm}
\headerdepth=0pt\hoffset=-23.81pt\voffset=0pt\hsize=499pt\vsize
=665.3pt\headline={\hfil}\footline={\fontsetaa\parindent=0pt
\leftskip=0pt\parfillskip=0pt plus1fil\rightskip=0pt\spaceskip
=0em\xspaceskip=0em
\noindent\setbox0=\hbox{{\it Space-Time Structure as Hidden
Variable}}\tabnone\setbox0=\hbox{{\it4/7/00}}\dimen1=-239.9pt
\tableft\setbox0=\hbox{}\dimen1=-39.98pt\tableft\setbox0=\hbox
{{\it page \folio\ of 9}}\dimen1=-39.98pt\tableft

\hfil}\footmargin=10pt\parskip=2pt plus1pt
minus.5pt\advance\belowdisplayshortskip by-\parskip\advance
\belowdisplayskip by-\parskip\raggedbottom\def\secskip{0pt}\fontsetab
\parindent=0pt\leftskip=0pt plus1fil\parfillskip=0pt\rightskip
=0pt plus1fil\spaceskip=0.3333em\xspaceskip=0.5em
{\bf Space-Time Structure as Hidden Variable}\par\nobreak

\fontsetaa\spaceskip=0.3333em\xspaceskip=0.5em
\relax Bart Jongejan

{\it C\ae ciliavej 31, 2500 Valby, Denmark}

{\it Electronic address: bart@cst.ku.dk}

\vskip0pt plus\secskip\penalty-250\vskip0pt plus-\secskip\vskip
11.4pt\fontsetaa\parindent=0pt\leftskip=18pt plus1fil\parfillskip
=0pt\rightskip=0pt plus1fil\spaceskip=0.3333em\xspaceskip=0.5em
\item{}\par\nobreak

\fontsetac\vskip 6pt\parindent=8.64pt\leftskip=54pt\parfillskip
=0pt plus1fil\rightskip=54.04pt\spaceskip=0em\xspaceskip=0em
\relax EPR correlations exist and can be observed independently
of any a priori given frame of reference. We can even construct
a frame of reference that is based on these correlations. This
observation-based frame of reference is equivalent to the customary
a priori given frame of reference of the laboratory when describing
real EPR experiments.

J.S. Bell has argued that local hidden parameter theories that
reproduce the predictions of Quantum Mechanics cannot exist,
but the counterfactual reasoning leading to Bell{\tt\char13}s
conclusion is physically meaningless if the frame of reference
that is based on EPR-correlations is accepted as the backdrop
for EPR-type experiments.

The refutal of Bell{\tt\char13}s proof opens up for the construction
of a viable hidden parameter theory. A model of a spin {\it
$\hbar$\/}/2 particle in terms of a non-flat metric of space-time
is shown to be able to reproduce the predictions of quantum
mechanics in the Bohm-Aharonov version of the EPR experiment,
without introducing non-locality.

\vskip 9.1pt\parindent=0pt
\relax PACS numbers: 03.65.Bz, 02.40.Ky, 04.20.Gz 

\vskip0pt plus\secskip\penalty-250\vskip0pt plus-\secskip\vskip
18pt\fontsetad\parindent=0pt\leftskip=18pt plus1fil\parfillskip
=0pt\rightskip=0pt plus1fil\spaceskip=0.3333em\xspaceskip=0.5em
\item{}\par\nobreak

\vskip0pt plus\secskip\penalty-250\vskip0pt plus-\secskip\vskip
6pt\fontsetac\parindent=0.4pt\spaceskip=0.3333em\xspaceskip
=0.5em
\item{I.}{\bf HIDDEN VARIABLES}\par\nobreak

\vskip0pt plus\secskip\penalty-250\vskip0pt plus-\secskip\vskip
10pt\parindent=0.8pt
\item{A.}{\bf The Einstein, Podolsky and Rosen }{\bi Gedanken}
{\bf experiment}\par\nobreak

\parindent=8.8pt\leftskip=0.8pt\parfillskip=0pt plus1fil\rightskip
=0pt\spaceskip=0em\xspaceskip=0em
\relax Albert Einstein was convinced that quantum mechanics
is an incomplete theory, which was a position opposite to that
of Niels Bohr. Their discussion culminated in a paper by Einstein,
Podolsky and Rosen (EPR) [1], in which they showed how one can
measure two non-commuting physical quantities to any degree
of accuracy. Bohr{\tt\char13}s reply [2] was soon to follow.
The issue is still debated.

The set-up of the EPR experiment consists of two observation
posts doing measurements of momentum or position on the flown-apart
members of particle pairs that have been carefully prepared
to have no net momentum relative to the laboratory frame.

The particles in a pair can be regarded as exact copies of each
other, apart from being mirrored: if the same measurement is
performed at both observation posts, then the outcomes are each
other's exact opposites. Given a quantity to be determined for
both particles, we can suffice with doing only one of the measurements.

EPR's idea was that two non-commuting quantities, such as momentum
and position, can be determined by measuring one of the quantities
directly and by deriving the value of the other quantity from
the outcome of the measurement of that quantity on the other
particle in the pair.

Bohm and Aharonov [3] devised a version of the EPR Gedanken
experiment that has been the focus of much theoretical and experimental
work. In their experiment, the measurements are done on pairs
of spin {\it$\hbar$\/}/2 particles that are prepared in the
singlet state, which is a quantum state that does not hold any
information about the directions of the spins of the individual
particles.

In the Bohm-Aharonov experiment the particles fly apart toward
two widely separated observation posts, where they traverse
the gaps of Stern-Gerlach magnets. During such a traversal,
due to a coupling between the particle's intrinsic spin and
the longitudinal gradient of the magnetic field, the particle's
path is bent either away from or toward the pole where the magnetic
field is strongest. The particle finally hits one of two detectors,
depending on which route it took. One of the detectors only
records particles that had spin up $\left(\uparrow\right)$,
while the other only records those with spin down $\left(\downarrow
\right)$, ``up'' and ``down'' being defined relative to the
Stern-Gerlach magnet. The detectors are fixed to their respective
Stern-Gerlach magnets, so that the directions that are ``up''
and ``down'' rotate together with the freely orientable Stern-Gerlach
magnets.

\vskip0pt plus\secskip\penalty-250\vskip0pt plus-\secskip\vskip
16pt\parindent=0.8pt\leftskip=18pt plus1fil\parfillskip=0pt
\rightskip=0pt plus1fil\spaceskip=0.3333em\xspaceskip=0.5em
\item{B.}{\bf Can counterfactual considerations complete the
description of physical reality?}\par\nobreak

\parindent=8.8pt\leftskip=0.8pt\parfillskip=0pt plus1fil\rightskip
=0pt\spaceskip=0em\xspaceskip=0em
\relax Employing propositions of the type that EPR used to show
that quantum mechanics can not be complete, J.S. Bell [4] showed
that any theory that reproduces the predictions made by quantum
mechanics and yet is more complete than quantum mechanics necessarily
postulates instantaneous action at a distance. In other words,
the kind of theories that Einstein envisaged as successors to
quantum mechanics would be difficult to reconcile with relativity
theory, which champions locality and does not allow any signal
to travel faster than light.

Bell's proof is dependent on counterfactual propositions. A
counterfactual proposition assigns a determinate value to a
quantity that could have been directly observable, but is not,
typically because another, incommensurable quantity is measured.
The experimental basis - if you can call it that - for this
assignment is the measurement of the same quantity on the far
away twin particle. The reasoning is that the measurement on
the twin is as good as the measurement on the particle itself,
because the inner states of the particles must be fully correlated
in order to preserve the isotropy of the quantum state of the
pair.

To give teeth to such a counterfactual proposition, not only
do we have to ascribe reality to the result of the measurement,
but also to the angle between the counterfactual orientation
of the instrument and the (factual or counterfactual) orientation
of the other instrument.

Let us hold EPR's and Bell's counterfactual reasoning against
the background of Riemannian geometry, or, more specifically,
general relativity. Only geometric relations, such as angles
and distances, between things that are local to each other in
space and in time bear physical meaning according to the general
relativity theory. Geometric relations over long spatio-temporal
distances, such as the angle between the directions of observation
in the Bohm-Aharonov experiment, are a different matter. 

If it is not assumed that space-time is flat, then only an operational
definition can lend physical meaning to these relations. In
general, different operational definitions can give physical
meaning to the same relation, but in theory the methods need
not agree on the outcomes of the measurements. This is clearly
exemplified by the multitude of theory laden operational definitions
for distances on cosmological scales.

What then is the angel between a factual set-up of a Stern-Gerlach
magnet at one of the two observation posts and a counterfactual
set-up of the Stern-Gerlach magnet at the other observation
post? Of course, because one of the two set-ups is not effectuated,
there is no direct means of measuring this angle. We can only
base our answer on interpolation, together with a smoothness
assumption that coordinates the interpolated but counterfactual
observation with real ones. Normally, interpolation depends
on a smoothness assumption that is innocuous, because the missing,
interpolated values may some day be replaced by outcomes of
real experiments. In such cases, interpolation is a falsifiable
theory and therefore acceptable. In Bell's proof, the interpolation
is {\it not\/} falsifiable, because the measuring apparatus
is sitting itself in the way. That weakens Bell's conclusion.
We can only accept Bell's proof if we assume that the space
time backdrop is smooth and constant enough to allow us to interpolate
between measurements, but this assumption excludes from consideration
any theory that denies that space-time is like that.

\vskip0pt plus\secskip\penalty-250\vskip0pt plus-\secskip\vskip
16pt\parindent=0.8pt\leftskip=18pt plus1fil\parfillskip=0pt
\rightskip=0pt plus1fil\spaceskip=0.3333em\xspaceskip=0.5em
\item{C.}{\bf The Bohm-Aharonov experiment without flat-space
preconception.}\par\nobreak

\parindent=8.8pt\leftskip=0.8pt\parfillskip=0pt plus1fil\rightskip
=0pt\spaceskip=0em\xspaceskip=0em
\relax Bell{\tt\char13}s proof hinges on the postulate that
space time is flat, but this postulate may be false. This is
the main theme of this paper and we will dwell on it a little
more, because understanding the epistemological restrictions
that relativity imposes is essential for appreciating the approach
towards hidden variables that is presented later.

First think of the Bohm-Aharonov experiment as a set-up consisting
of two observation posts connected by the floor of a laboratory
or something else that we may regard as rigid. Each observation
post consists of a Stern-Gerlach magnet that is freely orientable
in its mounting into any of a large number of directions, or
lines of observation. Each such orientation is identifiable,
for example by reading off the color of a mark on the mounting
that a pointer aligned with the axis of the magnet is pointing
at. There may be many differently colored marks, each identifying
a unique orientation of the magnet. If one wishes so one can
define the mountings of the instrument to include far away stars,
which then can be used as the marks for that post. We also have
calibrated scales on the mounting so that we have the option
to read off the angular coordinates of the direction vector
of the Stern-Gerlach magnet. A post{\tt\char13}s electron detectors
are fixed to the Stern-Gerlach magnet inside that post. The
rigid connection between the two posts (or rather: between the
mountings), together with conventional means of doing geodesy
(measuring rods, light beams, gyroscopes) provides us with a
reference frame in which both measuring instruments have definite
positions and orientations. Even counterfactual orientations
can be tracked, because the rigid frame {\tt"}fixes{\tt"} all
thinkable orientations. This is the normal, {\tt"}robust{\tt
"} experimental context of Bell{\tt\char13}s proof.

Now remove all unnecessary equipment: the measuring rods, light
beams, gyroscopes, scales and even the laboratory floor. Would
that make any difference? We did not do away the mountings of
the instruments and are therefore still able to identify each
orientation by reading off the color of the mark that a Stern-Gerlach
magnet is pointing at. Can we reconstruct the experiment with
this basic equipment?

From each pair of measurements we obtain a data-triplet: the
color of the mark that the left magnet was pointing at, idem
for the right magnet and finally the outcome of the detectors,
which arbitrarily may be defined to be {\tt"}S{\tt"} (for {\tt
"}same{\tt"}) if both {\tt"}up{\tt"} or both {\tt"}down{\tt
"} detectors were hit and {\tt"}N{\tt"} (for {\tt"}not same{\tt
"}) if one {\tt"}up{\tt"} and one {\tt"}down{\tt"} detector
was triggered. Our logbook will have just three columns, the
experiment has only three degrees of freedom.

After doing a long series of such experiments, with the magnets
having been oriented in all possible combinations of directions
many times, the log of outcomes will enable us to assign statistical
probabilities for measuring the same spin component to each
pair of orientations of the Stern-Gerlach magnets. We might
for example observe that {\tt"}yellow{\tt"} left and {\tt"}blue{\tt
"} right have a 73\% chance of resulting in the value {\tt"}S{\tt
"}. There is a conceptual difference with the full blown experimental
set-up, though: we have no prior knowledge of the geometric
relations between the Stern-Gerlach magnets; we can only pairwise
statistically relate marks on the mountings with each other.
We do not even know the angles between two different orientations
of the same magnet!

Now we will try to organize the three columns of obtained data
by mapping the colored marks that identify directions of observations
onto points on a sphere in such a way that exactly one point
is assigned to each mark. This mapping must map the marks on
both surroundings, however far separated from each other, onto
a single sphere that does not exist physically, but only as
a mathematical tool. Underlying the mapping is the working hypothesis
that there is a functional relation between the statistical
probability to measure the same component and the angle, as
measured on this sphere, between the orientations of the Stern-Gerlach
magnets. We can start to assume a linear dependency: a 100\%
probability and a 0 \% probability correspond to angles of $180{{}^\circ
}$ and of $0{{}^\circ}$ between the magnets, a 10\% probability
would correspond to $18{{}^\circ}$, and so on. Such a relation
would be appropriate if the spinning particles were macroscopic
objects with observationally well defined {\tt"}north{\tt"}
and {\tt"}south{\tt"} hemispheres. However, if we apply this
relation to our data then the obtained angles force us to map
the same mark onto several points, which is not what we wanted.

Eventually, we would find that
$$\displayindent=0.8pt\displaywidth=239.7pt P\left(S\right)=probability\,to\,measure\,same\,value=$$
$$\displayindent=0.8pt\displaywidth=239.7pt=\sin^2\left(angle\,between\,orientations\over
2\right)\eqno(1a)$$

establishes a 1-1 mapping, as does its mirroring twin
$$\displayindent=0.8pt\displaywidth=239.7pt P\left(S\right)=probability\,to\,measure\,same\,value=$$
$$\displayindent=0.8pt\displaywidth=239.7pt=\cos^2\left(angle\,between\,orientations\over
2\right).\eqno(1b)$$

Now, not only would we have learned how to define angles between
the two measuring instruments when directed towards two given
colored marks, we would also have a consistent way to figure
out the angles between the colored marks at one and the same
post. That means that we would have regained the spherical geometry
relations that we did not assume as given a priori. That, in
turn, would enable us to deliver exactly the same kind of experimental
report as someone who had a rigid frame to connect the measuring
posts and rods, light beams, gyroscopes and scales to measure
the geometry. We could also verify the contingent fact that
this statistical way of determining geometric relations is perfectly
consistent with other, more conventional means, such as with
the help of rods, light beams, gyroscopes and scales. But of
course, as little physical sense it makes to ascribe temperature
and pressure (or the mean kinetic energy and momentum per particle)
to a single particle in a gas, as little sense would it make
to ascribe the statistically defined angle to a pair of measurements
that contributed to the very determination of the angle. {\it
A fortiori}, we can not draw firm conclusions from an argument
that hinges on a counterfactual set-up controlled by the statistically
defined angle. There is no observational basis for assigning
a value to such an angle, because the value, as defined operationally
in the above way, is of statistical character and therefore
not applicable to any pair of orientations in any specific pair
of measurements, but only in the long run of many measurement
event pairs. In addition, the assignment of a value of the angle,
operationally defined as above, to pairs of orientations that
are not both effectuated, can only be based on an arbitrary
convention and therefore renders any argument that is based
on such angles unconvincing.

If the presented way of constructing a frame of reference is
so feeble that it does not allow us to assign values to angles
between counterfactual setups, why then should we not stick
with the conventional means using rods, gyroscopes and so on?
The reason is that our method is not any more feeble than conventional
means! Unless conventional methods mysteriously gain strength
somewhere in the transition from the quantum to the classical
regime, {\it any\/} angle that can be measured by conventional
means can also be measured using a sufficiently large ensemble
of spin component observations on an equally large number of
pairs of spin {\it$\hbar$\/}/2 particles, to any desired accuracy.
However, the proposed method, which is obviously based on quantum
phenomena, has the additional advantage that it very clearly
delineates the domain of applicability of the method; a domain
of applicability that can not be surpassed by conventional methods,
unless the aforementioned mysterious powers opened a back door
for the conventional methods to do measurements of angles that
are out of reach for the proposed method.

Whereas proofs like Bell{\tt\char13}s are hard pressed because
of the lack of an observational basis for the assumed frame
of reference, a realist point of view is not obstructed in the
same way. It is sensible to imagine a counterfactual set-up
of a measuring apparatus that is oriented towards a particular
colored mark and with a definite value of the spin component
in that direction, but we must keep in mind that the whereabouts
of the instrument{\tt\char13}s orientation relative to other
(factual or counterfactual) orientations is unknown. Underlying
Bell{\tt\char13}s and similar proofs is a concept of realism
that is far more encompassing than necessary. Things that derive
the status of being part of reality by force of real observations,
such as the angle between directions of observation and even
systems of coordinates in general, can not be idealized to an
existence detached from these observations without introducing
trouble in some corners of our theoretical picture of the world.

\vskip0pt plus\secskip\penalty-250\vskip0pt plus-\secskip\vskip
16pt\parindent=0.4pt\leftskip=18pt plus1fil\parfillskip=0pt
\rightskip=0pt plus1fil\spaceskip=0.3333em\xspaceskip=0.5em
\item{II.}{\bf MATHEMATICAL CONSTRAINTS ON HIDDEN VARIABLE THEORIES
OF SPIN }{\it$\hbar$\/}{\bf/2 PARTICLE}\par\nobreak

\vskip0pt plus\secskip\penalty-250\vskip0pt plus-\secskip\vskip
10pt\parindent=0.8pt
\item{A.}{\bf What makes an aspirant HV theory?}\par\nobreak

\parindent=8.8pt\leftskip=0.8pt\parfillskip=0pt plus1fil\rightskip
=0pt\spaceskip=0em\xspaceskip=0em
\relax We have seen that the Bohm-Aharonov experiment has just
three relevant degrees of freedom: color of left mark, color
of right mark and the combination of the outcomes. We will now
discuss hidden variable theories that also exhibit three degrees
of freedom and hope to find one that can be made to correspond
to the Bohm-Aharonov experiment and that explains the statistical
correlations found in the Bohm-Aharonov experiment (which are
assumed to be accurately predicted by quantum mechanics).

The contemplated hidden variable theories all have one aspect
in common: not only the Stern-Gerlach magnets have definite
orientations, also the particle itself has one, which is the
axis of rotational symmetry or {\tt"}spin axis{\tt"}. The three
variables that specify any configuration of the three directions
are ({\it a\/}) the angle between the left measuring apparatus
and the spin axis, ({\it b\/}) the angle between the right measuring
apparatus and the spin axis and ({\it c\/}) the angle between
the left measuring apparatus and the right measuring apparatus.
None of these variables are precisely measurable, but each corresponds
to observed data: if the left {\tt"}up{\tt"} detector is triggered,
then the angle between the left measuring apparatus and the
spin axis is less than $90{{}^\circ}$. If the {\tt"}down{\tt
"} detector is triggered, the angle is somewhere between $90{{}^\circ
}$ and $180{{}^\circ}$. Likewise for the right detector. The
angle between the measuring apparatuses is taken to be the angle
that was statistically derived from the outcomes of a long series
of Bohm-Aharonov measurements, using Eq. (1b).

The main point made by Bell was that no {\it local\/} hidden
variable theory is able to reproduce the predictions of quantum
mechanics. The requirement that our aspirant hidden variable
theories are local puts three constraints on the statistical
distribution of the angles between the measuring apparatuses
and between each measuring apparatus and the particle{\tt\char
13}s spin axis. These will be discussed now.

\vskip0pt plus\secskip\penalty-250\vskip0pt plus-\secskip\vskip
9.1pt\leftskip=19.2pt
\relax Constraint 1. {\it\ The orientations of the measuring
instruments are unrelated.}\par\nobreak

\leftskip=1.2pt
\relax Each measuring apparatus is oriented in a way that does
not depend on the orientation of the other instrument, not even
statistically. If the movements of the measuring apparatuses
A and B were restricted to a plane, then this constraint would
translate to a uniform distribution of angles $\angle AB$ in
the range $0\leq\angle AB\leq\pi$. We do, however, assume that
the instruments are freely orientable in space. In that case,
the probability that the angle between A and B is $\angle AB$
is proportional to $\sin\angle AB$.

Whereas the angle is non-uniformly distributed, its cosine is
not. So we require a uniform distribution $\rho\left(Z_{AB}\right
)$ of the inner product $Z_{AB}={\rm-cos}\angle AB=-{\bi a}.{\bi
b}$ in the range $-1\leq Z_{AB}\leq1$. The minus sign is arbitrarily
introduced to compensate for the circumstance that the particles{\tt
\char13} spins are {\it anti\/}-parallel.

Normalization requires $\int^1_{-1}\rho\left(Z_{AB}\right)dZ_{AB}=1$,
so that 
$$\displayindent=1.2pt\displaywidth=239.3pt\rho\left(Z_{AB}\right
)=\onehalf\eqno(2)$$

It is worthwhile to indicate which role the set-up plays. We
ask that during each measurement the measuring apparatus points
at a randomly chosen point of its own surroundings. This does
not automatically imply an isotropic distribution of the orientations
with respect to each other: one could imagine that each observation
post{\tt\char13}s orientations, taken separately, would survive
a {\tt"}randomness test{\tt"}, but that the orientations were
not distributed isotropically with respect to each other. That
situation could arise if the observers used the same sequence
of random numbers to prepare the instruments for each pair of
measurements. We assume that such correlation does not occur,
because that seems to be the only assumption that is compatible
with the principles of locality, causality and free will. 

\vskip0pt plus\secskip\penalty-250\vskip0pt plus-\secskip\vskip
16pt\leftskip=19.2pt
\relax Constraint 2. {\it(Locality condition.) The orientation
of one magnet does not influence the result obtained with the
other.}\par\nobreak

\leftskip=1.2pt
\relax Suppose that someone came up with a HV theory of a spin
{\it$\hbar$\/}/2 particle. In order to test the claim that it
reproduces the predictions of QM in Bohm-Aharonov experiments,
we had to subject the theory to a Gedanken experiment in which
a great number of spin-component measurements were randomly
chosen. How would the randomly chosen orientations of a measuring
instrument be distributed with respect to the particle{\tt\char
13}s axis of rotational symmetry? As we have no means of observing
this distribution, we postulate one. In the absence of any reason
to assume a non-isotropic distribution, we assume the isotropic
distribution. Call the inner product of the orientation of the
instrument and the direction of the particle (denoted by unit
vectors) $Z_A$ and $Z_B$ (for instrument A and instrument B).
Require that $\rho\left(Z_A\right)=\rho\left(Z_B\right)=\onehalf
$.

The sign of $Z_A$ determines the outcome of the measurement
made with the magnet at observation post A. The locality condition
says that $Z_A$ is independent of the orientation of instrument
B. The angle between the instruments can be taken to represent
this orientation, in which case we locality condition translates
to
$$\displayindent=1.2pt\displaywidth=239.3pt\rho\left(Z_A,Z_{AB}\right
)=\rho\left(Z_A\right)\rho\left(Z_{AB}\right)=1/4.\eqno(3a)$$

Alternatively, we can take the (hidden) angle between instrument
B and the spin axis as representing the orientation, so we also
require that
$$\displayindent=1.2pt\displaywidth=239.3pt\rho\left(Z_A,Z_B\right
)=\rho\left(Z_A\right)\rho\left(Z_B\right)=1/4.\eqno(3b)$$

As we are used to specify orientations with two mutually independent
angles, it is tempting to require that $Z_A$ is independent
of both of $Z_{AB}$ and $Z_B$:
$$\displayindent=1.2pt\displaywidth=239.3pt\rho\left(Z_A,Z_B,Z_{AB}\right
)=\rho\left(Z_A\right)\rho\left(Z_B\right)\rho\left(Z_{AB}\right
)=1/8.\eqno(4)$$

However, below we will see that this conflicts with the next
constraint and even with the statistics of classical spinning
particles.

\vskip0pt plus\secskip\penalty-250\vskip0pt plus-\secskip\vskip
16pt\leftskip=19.2pt
\relax Constraint 3. {\it The theory reproduces the predictions
of quantum mechanics}\par\nobreak

\leftskip=1.2pt
\relax The correlation between the outcomes of measurements
on flown-apart particles with counter parallel spin axis conforms
exactly to the predictions of QM (see also Eq. (1b)):
$$\displayindent=1.2pt\displaywidth=239.3pt P\left(S\right)\equiv
P\left(\uparrow\uparrow\right)+P\left(\downarrow\downarrow\right
)=\sin^2{\angle AB\over2}$$
$$\displayindent=1.2pt\displaywidth=239.3pt={1+Z_{AB}\over2}\eqno
(5a)$$
$$\displayindent=1.2pt\displaywidth=239.3pt P\left(N\right)\equiv
P\left(\uparrow\downarrow\right)+P\left(\downarrow\uparrow\right
)=\cos^2{\angle AB\over2}$$
$$\displayindent=1.2pt\displaywidth=239.3pt={1-Z_{AB}\over2}.\eqno
(5b)$$

\vskip0pt plus\secskip\penalty-250\vskip0pt plus-\secskip\vskip
16pt\parindent=0.8pt\leftskip=18pt plus1fil\parfillskip=0pt
\rightskip=0pt plus1fil\spaceskip=0.3333em\xspaceskip=0.5em
\item{B.}{\bf The joint distribution }$\rho\left({\bi Z}_{\bi
A},{\bi Z}_{\bi B},{\bi Z}_{{\bi A}{\bi B}}\right)${\bf\ that
explains the quantum mechanical predictions.}\par\nobreak

\parindent=8.8pt\leftskip=0.8pt\parfillskip=0pt plus1fil\rightskip
=0pt\spaceskip=0em\xspaceskip=0em
\relax We assume that in a HV theory, a full specification of
a measurement of two spin components in the Bohm-Aharonov experiment
requires three angles, or there cosines. We must now investigate
whether there are distributions $\rho\left(Z_A,Z_B,Z_{AB}\right
)$ of these three quantities that fulfill all three constraints.
For example, the first constraint requires that
$$\displayindent=0.8pt\displaywidth=239.7pt\rho\left(Z_{AB}\right
)=\int^1_{-1}\int^1_{-1}\rho\left(Z_A,Z_B,Z_{AB}\right)dZ_AdZ_B=1/2,\eqno
(6)$$

and second constraint, the locality condition, requires that
$$\displayindent=0.8pt\displaywidth=239.7pt\rho\left({Z_A,Z}_{AB}\right
)=\int^1_{-1}\mathop{}_{}\rho\left(Z_A,Z_B,Z_{AB}\right)dZ_B=1/4.\eqno
(7)$$

Finally, according to quantum mechanics we must find that
$$\displayindent=0.8pt\displaywidth=239.7pt P\left(\uparrow
\uparrow\right)=\rho{\left(Z_{AB}\right)}^{-1}\int^1_0\int^1_0\rho
\left(Z_A,Z_B,Z_{AB}\right)dZ_AdZ_B$$
$$\displayindent=0.8pt\displaywidth=239.7pt={1+Z_{AB}\over4}.\eqno
(8)$$

As it did not seem a trivial task to solve the set of equations
constituting the three constraints, a computer aided approach
was chosen. It was not difficult to find a distribution that
fulfills constraints 1 an 2, Eq. (4) is such a distribution.
Then, repeatedly applying an algorithm that transforms the distribution
to a new distribution that also fulfills constraints 1 and 2,
distributions were found that fulfill the third constraint as
well. The algorithm is as follows: Choose two values for each
of $Z_A,\,Z_B$ and $Z_{AB}$. These values are the coordinates
of eight cells, the probability density of which we are going
to redistribute. Choose an amount $\triangle\rho$ and add this
amount to four cells spanning a tetrahedron and subtract the
same amount from the remaining cells. By the right choice of
$\triangle\rho$ we can empty at least one of the eight cells.
See Fig. 1. Some of the results of this discrete approximation
can be seen in Figs. 2-4.

The distribution in Eq. (4) fully acknowledges the freedom of
the experimenters to vary the angle between the instruments
($Z_{AB}$) and it guarantees the isotropic distribution of the
axis of the model relative to the lines of observation ($Z_A$
and $Z_B$). Yet this distribution is not realistic at all, because
it does not restrict the angles between the instruments ($\arccos
{-Z}_{AB}$) and the angles between the measurement instruments
and the axis of the model ($\arccos Z_A$ and $\arccos Z_B$).
These three angles can not be completely independent: for example
can the angle between the measuring instruments not exceed the
sum of the angles between the instruments and the axis of the
model.

If two of the angles already are fixed to any values between
$0$ and $\pi$ (any two of $\{\arccos Z_{AB}$,$\arccos Z_A$,$\arccos
Z_B\}$, call them $\alpha_1$ and $\alpha_2$), then we have the
following constraint on the third angle $\alpha_3$:
$$\displayindent=0.8pt\displaywidth=239.7pt\left|\alpha_1-\alpha
_2\right|\leq\alpha_3\leq\alpha_1+\alpha_2.\eqno(9)$$

That means that whereas any two of the three angles are independent
of each other, there exists a mutual dependency between the
three angles.

The uniform distribution of three vectors ${\bi a},{\bi b}$
and $\bi c$ over all directions illustrates this dependency.
In an Euclidean frame of reference, the joint density of the
three inner products $s=Z_A={\bi a}.{\bi c},\,t=Z_B={\bi b}.{\bi
c},\,$$u=Z_{AB}={\bi a}.{\bi b}$ is
$$\displayindent=0.8pt\displaywidth=239.7pt\rho\left(s,t,u\right)={\left(8\pi\sqrt{1+2stu-s^2-t^2-u^2}\right
)}^{-1}$$
$$\displayindent=0.8pt\displaywidth=239.7pt\qquad\qquad\left(1+2stu-s^2-t^2-u^2>0\right)$$
$$\displayindent=0.8pt\displaywidth=239.7pt=0\qquad(1+2stu-s^2-t^2-u^2\leq0).\eqno(10)$$

Figure 3 illustrates the mutual dependency between the three
directions and the remarkable discontinuity of the density at
the border between possible and impossible configurations, 

The next step is the fulfillment of constraint 3. Using the
transformation algorithm again, we eventually approach a distribution
that reproduces the predictions of quantum mechanics (Eq. 5a/b).
There are many distributions that come very close, but they
are all characterized by regions of almost emptiness and steep
climbs to high values of probability density. The following
distribution, which is the limiting distribution as the number
of subdivisions in the discrete approximation goes to infinity,
and which uses Dirac delta functions, fulfills all three constraints
(see Fig. 4 for a discrete approximation):
$$\displayindent=0.8pt\displaywidth=239.7pt\rho\left(Z_A,Z_B,Z_{AB}\right
)=1/8\left[\delta\left(Z_A+Z_B+Z_{AB}+1\right)\right.$$
$$\displayindent=0.8pt\displaywidth=239.7pt+\,\delta\left(Z_A-Z_B-Z_{AB}+1\right
)$$
$$\displayindent=0.8pt\displaywidth=239.7pt+\,\delta\left(Z_A+Z_B-Z_{AB}-1\right
)$$
$$\displayindent=0.8pt\displaywidth=239.7pt\left.+\,\delta\left
(Z_A-Z_B+Z_{AB}-1\right)\right]\eqno(11)$$

This distribution indicates that one continuous degree of freedom
is replaced by a discrete one: the density is only non-zero
on the surface of a tetrahedron spanned by four of the eight
corners of the configuration cube.

A classical configuration of three independent vectors is specified
by three numbers, which are the lengths of the sides of a triangle
on the unit sphere. They can not live within less than the two
dimensions of this sphere. On the other hand, configurations
that are compatible with QM require only two numbers and a sign,
the third number being a function of either the sum or the difference
of the other two, depending on the sign.

Perhaps somewhat unexpectedly, our search for a probability
distribution for the angle between the instruments and the angles
between each instrument and the (hidden) spin axis did not merely
result in a non-classical distribution, but also in a qualitative
characterization of any hidden variable theory with hopes to
fulfill all three constraints: the theory must endow a model
of a spin {\it$\hbar$\/}/2 particle with a degeneracy that replaces
one continuous degree of freedom with a two-valued one. We will
now look at a theory that accomplishes this feat.

\vskip0pt plus\secskip\penalty-250\vskip0pt plus-\secskip\vskip
16pt\parindent=0.4pt\leftskip=18pt plus1fil\parfillskip=0pt
\rightskip=0pt plus1fil\spaceskip=0.3333em\xspaceskip=0.5em
\item{III.}{\bf EXAMPLE: A HIDDEN VARIABLE MODEL BASED ON NON-FLAT
SPACE-TIME STRUCTURE}\par\nobreak

\vskip0pt plus\secskip\penalty-250\vskip0pt plus-\secskip\vskip
10pt\parindent=0.8pt
\item{A.}{\bf Overview}\par\nobreak

\parindent=8.8pt\leftskip=0.8pt\parfillskip=0pt plus1fil\rightskip
=0pt\spaceskip=0em\xspaceskip=0em
\relax In the foregoing, a weakness in Bell{\tt\char13}s counterfactual
reasoning was exposed by peeling away unwarranted and mostly
silent assumptions that are underlying his proof, until we were
left with the nitty gritty observational data. Then we postulated,
in spite of Bell{\tt\char13}s conclusion, that the direction
of a particle{\tt\char13}s spin is really existing, although
hidden from observation. We found that the QM-compatible configurations
of three vectors, denoting the orientations of the measuring
instruments and a candidate model{\tt\char13}s hidden variable,
seem to live within one dimension less than classically expected.

The next step is the construction of a model with a characteristic
that we rightly can call the spin direction of the model. We
postulate that the stuff that the particle is made of is space
time itself, or more specifically, the structure of space time.
We will check that the full specification of a geodetic path
herein - tentatively representing the world line of a measuring
apparatus during a measurement - exhibits the same lack of one
degree of freedom.

Such a postulated space-time structure is in part based on guesswork,
in part on esthetic rules, such as simplicity and symmetry.
The real structure of space time is perhaps unknowable, but
we may hit upon a theory of the structure of space time that
survives the observations that we perform to test it.

The presented model tries only to explain a very limited set
of phenomena, namely the correlation between two spin component
measurements. We have not tried very hard to incorporate and
explain other phenomena. Thus, a simple thing like the spatial
distance between the spinning particles is not expressed very
well by the proposed model, nor their relative movements. In
fact, the model explains two widely different phenomena without
making a distinction, which shows that in any case the differences
between these phenomena are not expressed in the model. The
first phenomenon is the correlation between the measurements
of spin components on two different particles that together
form a system in the singlet state. The second phenomenon, that
is explained equally well, is the passage of a single spinning
particle first through one, then through a second Stern-Gerlach
magnet at some distance from the first, that is inclined with
respect to the first.

\vskip0pt plus\secskip\penalty-250\vskip0pt plus-\secskip\vskip
16pt\parindent=0.8pt\leftskip=18pt plus1fil\parfillskip=0pt
\rightskip=0pt plus1fil\spaceskip=0.3333em\xspaceskip=0.5em
\item{B.}{\bf Metric and geodesic equations}\par\nobreak

\parindent=8.8pt\leftskip=0.8pt\parfillskip=0pt plus1fil\rightskip
=0pt\spaceskip=0em\xspaceskip=0em
\relax Consider a metric $g_{ik}$
$$\displayindent=0.8pt\displaywidth=239.7pt ds^2=\cos^2\vartheta
\,dt^2-dr^2-r^2d\vartheta^2-r^2\sin^2\vartheta\,d\varphi^2$$
$$\displayindent=0.8pt\displaywidth=239.7pt+\,2r\sin^2\vartheta
\,d\varphi dt.\eqno(12)$$

To get a feeling of how it would be like to be in a space time
with this metric, imagine an infinite set of concentric spheres.
Suppose that you momentarily were attached to one of these spheres.
You would have the feeling that this sphere was rotating around
the $\vartheta=0$ axis with angular velocity ${1\over r}={1\over
radius\,of\,sphere}$. In other words, you would feel an acceleration
away from the $\vartheta=0$ axis. If you were to move in the
direction opposite to this intrinsic rotation with angular velocity
$-1/r$ the acceleration would disappear. Although everywhere
the experienced acceleration could be explained as the effect
of an angular velocity that gradually decreased as $1\over r$,
the spheres in this space time are fixed to each other forever.
It is the curvature of space-time that gives the experience
of being accelerated, just like the curvature of space-time
caused by the Earth{\tt\char13}s mass lets us feel the force
of gravitation: an acceleration without movement, as opposed
to the acceleration experienced in a rocket.

The rotational effects in this space-time are due to the last
term in the metric ground form, $+2r\sin^2\vartheta\,d\varphi
d\vartheta$. This term specifies the model{\tt\char13}s spin
direction relative to the chosen frame of reference; by changing
the sign of this term or by rotating $180^\circ$ along an axis
perpendicular to the $\vartheta=0$ axis we obtain a model with
opposite spin direction.

In order to investigate how free test-particles move in this
space-time we have to solve the equations of motion
$$\displayindent=0.8pt\displaywidth=239.7pt{d^2x_i\over ds^2}=-\Gamma
^i_{jk}{dx_j\over ds}{dx_k\over ds}.\eqno(13)$$

The non-zero coefficients of the affine connection $\Gamma^i_{jk}$,
which are symmetric in the lower indices, are:

\vskip 2pt\leftskip=0pt\noindent\vbox{\tabskip=0pt\normalspacing
\offinterlineskip\halign to 239.7pt{#\tabskip=\twtabskip&#\strut
\hfil&#&#\strut\hfil&#\tabskip=0pt\cr
&$\Gamma^t_{tr}={\sin^2\vartheta\over2r}$&&\fontsetaa\fontsetac
$\Gamma^t_{r\varphi}=-\,{\sin^2\vartheta\over2}$&\cr
&\fontsetaa\fontsetac$\Gamma^r_{t\varphi}=_{}{\sin^2\vartheta
\over2}$&&\fontsetaa\fontsetac$\Gamma^r_{\vartheta\vartheta
}=-\,r$&\cr
&\fontsetaa\fontsetac$\Gamma^r_{\varphi\varphi}=-\,r\sin^2\vartheta
$&&\fontsetaa\fontsetac$\Gamma^\vartheta_{t\varphi}=\,{\sin
\vartheta\cos\vartheta\over r}$&\cr
&\fontsetaa\fontsetac$\Gamma^\vartheta_{r\vartheta}={1\over
r}$&&\fontsetaa\fontsetac$\Gamma^\vartheta_{\varphi\varphi}=-\sin
\vartheta\cos\vartheta$&\cr
&\fontsetaa\fontsetac$\Gamma^\vartheta_{tt}=-\,{\sin\vartheta
\cos\vartheta\over r^2}$&&\fontsetaa\fontsetac$\Gamma^\varphi
_{tr}=-\,{\cos^2\vartheta\over2r^2}$&\cr
&\fontsetaa\fontsetac$\Gamma^\varphi_{r\varphi}={1+\cos^2\vartheta
\over2r}$&&\fontsetaa\fontsetac$\Gamma^\varphi_{\vartheta\varphi
}={\cos\vartheta\over\sin\vartheta}${\fontaaa.}&\cr
}}
$$\displayindent=0.8pt\displaywidth=239.7pt\eqno\left(14\right
)$$

\leftskip=0.8pt\spaceskip=0em\xspaceskip=0em
\relax With the shorthand notation $U^r=r'={dr\over ds}$, $U^{r'}=r''={d^2r\over
ds^2}$ etc. and the above connection $\Gamma^i_{jk}$ the geodesic
equations become
$$\displayindent=0.8pt\displaywidth=239.7pt {U^t}'={-\,U^r\sin\vartheta\left[\sin\vartheta
(U^t-rU^\varphi)\right]\over r}\eqno
(15a)$$
$$\displayindent=0.8pt\displaywidth=239.7pt {U^r}'={(rU^\vartheta)^2-({rU^\varphi\sin}_{}\vartheta)[\sin
\vartheta(U^t-rU^\varphi)]\over r}\eqno
(15b)$$
$$\displayindent=0.8pt\displaywidth=239.7pt {U^\vartheta}'={-\,2\,U^r(rU^\vartheta)+{\cos\vartheta\over
\sin\vartheta}[\sin\vartheta(U^t-rU^\varphi)]^2\over r^2}\eqno
(15c)$$
$$\displayindent=0.8pt\displaywidth=239.7pt {U^\varphi}'={1\over r^2\sin\vartheta}
\times\bigl\{-U^r(rU^\varphi\sin\vartheta)$$
$$\displayindent=0.8pt\displaywidth=239.7pt+\,U^r\cos^2\vartheta
\left[\sin\vartheta(U^t-rU^\varphi)\right]$$
$$\displayindent=0.8pt\displaywidth=239.7pt+\,2{\cos\vartheta
\over\sin\vartheta}(rU^\vartheta)\left[\sin\vartheta(U^t-rU^\varphi
)\right]\bigr\}.\eqno(15d)$$

The geodesic equations have the following solutions:
$$\displayindent=0.8pt\displaywidth=239.7pt U^t=P+{X\over r}\eqno
(16a)$$
$$\displayindent=0.8pt\displaywidth=239.7pt U^\varphi={P-{X{\rm
cotg}^2\vartheta\over r}\over r}\eqno(16b)$$
$$\displayindent=0.8pt\displaywidth=239.7pt{\left(U^\vartheta
\right)}^2={A-{X^2\over\sin^2\vartheta}\over r^4}\eqno(16c)$$
$$\displayindent=0.8pt\displaywidth=239.7pt{\left(U^r\right
)}^2=-\,{A-X^2\over r^2}+{2PX\over r}+P^2-W,\eqno(16d)$$

where $P,X,A$ and $W$ are constants. We can also write
$$\displayindent=0.8pt\displaywidth=239.7pt P=\cos^2\vartheta
\,U^t+r\sin^2\vartheta\,U^\varphi\eqno(17a)$$
$$\displayindent=0.8pt\displaywidth=239.7pt X=r\sin^2\vartheta
\left(U^t-rU^\varphi\right)\eqno(17b)$$
$$\displayindent=0.8pt\displaywidth=239.7pt A={\left(r^2U^\vartheta
\right)}^2+{\left[r\sin\vartheta\left(U^t-rU^\varphi\right)\right
]}^2\eqno(17c)$$
$$\displayindent=0.8pt\displaywidth=239.7pt W={\left(\cos\vartheta
\,U^t\right)}^2-{\left(U^r\right)}^2-{\left(rU^\vartheta\right
)}^2-{\left(r\sin\vartheta\,U^\varphi\right)}^2$$
$$\displayindent=0.8pt\displaywidth=239.7pt+\,2r\sin^2\vartheta
\,U^tU^\varphi\eqno(17d)$$

$W$ is simply $g_{ij}U^iU^j$, the length of the vector $\bi
U$ squared. Time-like geodesics have $W>0$, space-like geodesics
have $W<0$ and light-like geodesics have $W=0$. We will restrict
$W$ to the values -1, 0 and 1. This restriction removes an arbitrary
scaling factor.

\vskip0pt plus\secskip\penalty-250\vskip0pt plus-\secskip\vskip
16pt\parindent=0.8pt\leftskip=18pt plus1fil\parfillskip=0pt
\rightskip=0pt plus1fil\spaceskip=0.3333em\xspaceskip=0.5em
\item{C.}{\bf Comparison with paths in central force fields}\par
\nobreak

\parindent=8.8pt\leftskip=0.8pt\parfillskip=0pt plus1fil\rightskip
=0pt\spaceskip=0em\xspaceskip=0em
\relax If we only look at time-like geodesics in the direction
from past to future (the paths that test particles follow, $W=1,\,U^t>0$),
then only three numbers $P,X$ and $A$ are needed to fully specify
a geodesic. What are the consequences of this paucity of orbit-fixing
numbers? Let us compare a geodesic in a central gravitational
field and a geodesic in this model-space time. Four constants
are needed to specify the orbit of a freely falling test particle
{\it A\/} with respect to a massive body, such as the orbit
of a planet around a star. The shape of the orbit or eccentricity
- whether the orbit is a circle, an ellipse, a parabola or a
hyperbola - provides one number. The size of the orbit - e.g.
the distance of closest approach to the central massive body
- provides another number. The orientation of the orbital plane,
defined as the unit vector normal to the orbital plane, requires
the specification of two angular coordinates. That adds two
more degrees of freedom and brings the total number of constants
to four.

For the time-like geodesics in our model-space-time the situation
is different. If we assume that two numbers are needed to specify
{\tt"}shape{\tt"} and {\tt"}size{\tt"}, then only one number
is left to specify the {\tt"}orbital plane{\tt"}. The quotes
indicate that we cannot be sure that it makes sense to talk
about shape, size and orbital plane. We will later have to look
at that.

If one number fixes an orbital plane then obviously the orbital
orientation has only one degree of freedom. Given one orbital
plane, then another orbital plane can be specified with reference
to the given orbital plane by providing the difference of the
{\tt"}orbital plane numbers{\tt"}. In fact, there is an ambivalence
in such a specification because of the sign of the difference.
This sign can not be specified without breaking the symmetry
between the two planes: we have to assume that either plane
can play the role of reference plane and the expressions should
not depend on this choice in any arbitrary way. 

We have seen the same {\tt"}directional degeneracy{\tt"} before:
the distribution of values $Z_A,Z_B,Z_{AB}$ that reproduces
quantum mechanics is such that given one $Z$-value, the other
$Z$-values can be specified with reference to this single $Z$-value.
For example, if $Z_A$ is given then $Z_B$ is specified, up to
a bivalent choice, by giving just one more number: $Z_{AB}$
(see 7a-d).

\vskip0pt plus\secskip\penalty-250\vskip0pt plus-\secskip\vskip
16pt\parindent=0.8pt\leftskip=18pt plus1fil\parfillskip=0pt
\rightskip=0pt plus1fil\spaceskip=0.3333em\xspaceskip=0.5em
\item{D.}{\bf How the constants define shape and orientation
of the geodesic}\par\nobreak

\parindent=8.8pt\leftskip=0.8pt\parfillskip=0pt plus1fil\rightskip
=0pt\spaceskip=0em\xspaceskip=0em
\relax We will now look more closely at the constants of the
motion and see whether it really is the case that there is only
one number to specify the orbital plane.

$A,P,X$ and $W$ are real numbers that only to some degree can
be chosen freely. The expression for ${\left(U^\vartheta\right
)}^2$ indicates that only values $\left|X\right|\leq\sqrt A$
can lead to geodesics. It also shows that such geodesics are
restricted to points where $\sin\vartheta\geq{\left|X\right
|\over\sqrt A}$. Geodesics with values of $\left|X\right|$ close
to $\sqrt A$ are close to the equatorial plane $\vartheta={\pi
\over2}$, while a value of $\left|X\right|\over\sqrt A$ close
to zero allows the geodesic to approach the poles very closely.
Therefore we define
$$\displayindent=0.8pt\displaywidth=239.7pt S={X\over\sqrt A}\qquad
\left(-1\leq S\leq1\right)\qquad$$
$$\displayindent=0.8pt\displaywidth=239.7pt''tilt\,of\,orbital\,plane''.\eqno
(18)$$

If $S=-1$ or $S=1$ then $\sin\vartheta=1$; such geodesics are
equatorial. All other geodesics are wavering north and south
(and through) the equatorial plane and their orbital plane,
if such a mathematical object can be defined, is tilted with
respect to the equatorial plane. The maximum angular distance
from the equatorial plane is reached when $\left|U^\vartheta
\right|=0$, which is when $\sin\vartheta=\left|S\right|$. We
call $S$ the {\tt"}tilt of the orbital plane with respect to
the equatorial plane{\tt"}, in analogy to the tilt of a planetary
orbit with respect to the Sun{\tt\char13}s equatorial plane,
which also is equal to the angle where the planet reaches its
greatest angular distance from the equatorial plane. While we
already have seen that the tilt of a planetary orbit is only
one of two constants that define the orientation of the orbital
plane, we still have to see whether there is such a second constant
in the case of our model-space-time geodesics.

We can learn much about the orbital shape and size from investigating
the radial velocity. If $A\ne X^2$ then ${\left(U^r\right)}^2$
is a second order function of $1/r$ and otherwise it is a first
order function of $1/r$. The coefficient of the $1/r^2$-term
is zero or negative, because $A\geq X^2$. From that follows
that ${\left(U^r\right)}^2$ has a maximum if $A\ne X^2$. If
the equation ${\left(U^r\right)}^2=0$has no solutions, then
this maximum is negative, which is forbidden, ${\left(U^r\right
)}^2$ being the square of a real number and therefore necessarily
non-negative. We can investigate the constraints on $P,X,W$
and $A$ that ensure that
$$\displayindent=0.8pt\displaywidth=239.7pt{\left(U^r\right
)}^2=-\,{A-X^2\over r^2}+{2PX\over r}+P^2-W=0\eqno(19)$$

has solutions.

The solutions of $1/r$ are:
$$\displayindent=0.8pt\displaywidth=239.7pt{1\over r}|_{U^r=0}={PX\pm
\sqrt{AP^2-AW+X^2W}\over A-X^2}.\eqno(20)$$

The condition that there be solutions is that
$$\displayindent=0.8pt\displaywidth=239.7pt P^2\geq W\left(1-{X^2\over
A}\right).\eqno(21)$$

The factor in parentheses is non-negative, because $A\geq X^2$.
If $W\leq0$, then the relation is fulfilled for all values of
$P$. For positive $W$ the above relation puts a lower bound
on the absolute value of $P$.

Because the radial parameter $r$ is non-negative, we are only
interested in non-negative solutions of $1/r$. If there is one
positive solution, then the trajectory is unbound: the positive
solution is the point of closest approach, but there is no point
of greatest radial distance. If the other solution is zero,
then the trajectory is just barely unbound and has the status
of a parabolic trajectory in a central force field. If the other
solution is negative, the trajectory is {\tt"}hyperbolic{\tt
"}. If there are two positive solutions, the trajectory is bound,
like the elliptic trajectories in a central field with a $1/r$
potential. If the two positive solutions are equal, the trajectory
has constant radius, i.e. it is comparable with circular orbits.

The close analogy between classical orbits in central force
fields and geodesics in our space-time model is concisely expressed
by 
$$\displayindent=0.8pt\displaywidth=239.7pt{P^2-W\over2}={{\left
(U^r\right)}^2+{\left(U^\perp\right)}^2\over2}-{XP\over r}-{X^2\over
{2r}^2},\eqno(22a)$$

where
$$\displayindent=0.8pt\displaywidth=239.7pt{\left(U^\perp\right
)}^2={\left(rU^\vartheta\right)}^2+\sin^2\vartheta(U^t-rU^\varphi
)^2\qquad$$
$$\displayindent=0.8pt\displaywidth=239.7pt''square\,of\,tangential\,velocity''.\eqno
(22b)$$

Eq. (22a) unmistakably has the signature of an energy, having
a kinetic part depending on the squares of the radial and tangential
velocities and two potential parts, one of which is due to a
long range $1/r$ potential that can be attractive or repelling,
like a Coulomb potential, while the other is due to a short
range attractive $1/r^2$ potential. The $1/r$ potential gives
rise to orbits having circular, elliptic, parabolic or hyperbolic
shapes, while the $1/r^2$ potential adds a precession of pericentrum
to the movement, like the shift of the perihelion of Mercury
that also is caused by a $1/r^2$ term.

The energy expression [Eq. (22a)] contains two independent constants
$P$ and $X$ that together define shape and size of an orbit
in the same way that shape and size of planetary orbits are
defined by the mass of the sun and the energy per kilogram of
planetary mass, where $X$ plays the role of the solar mass and
$\left(P^2-W\right)/2$ the role of energy per unit of planetary
mass.

Now we have {\tt"}used{\tt"} three constants to specify the
tilt of the orbital plane [Eq. (18)], the size of the orbit
[Eq. (20)] and the shape of the orbit [Eq. (22a)] and there
is no constant to completely specify the orientation$ $ of the
orbital plane. This situation is explained by the fact that
the orbital plane has to rotate to ensure that a geodesic test
particle does not leave the orbital plane. The pace $\Omega
={d\psi\over dt}$ with which the intersection line $\left\{\vartheta
=\pi/2,\varphi=\psi\left(t\right)\right\}$ of orbital plane
and the equatorial plane rotates depends on the instantaneous
radial distance $r$ of the test particle:
$$\displayindent=0.8pt\displaywidth=239.7pt\Omega=1/r.\eqno
(23)$$

The movement of the test particle for the case where the orbit
has constant $r$ ({\tt"}circular{\tt"} orbit) is accurately
modeled by the movement of a point on the rim of a coin that
is set to spin on its side on a table. The coin may start almost
upright, slowly falling due to frictional dissipation and decreasing
its tilt with respect to the surface of the table until it lies
down flat on the table. In the model space-time, of course,
there is no friction and the initial tilt remains the same forever.
While the coin is wobbling on the table, it rolls on its surface,
which means that points on the rim not only take part in the
rotation of the plane of the coin, but also in a rotation around
the axis that is perpendicular to the coin, moving up and down
and around in a complex dance.

The movement becomes even more complex if the radial distance
is not constant, but it can still be understood easily if one
imagines that the movement takes place in a tilted orbital plane
that rotates. Figures 5-6 give depict a geodesic from these
two perspectives.

\vskip0pt plus\secskip\penalty-250\vskip0pt plus-\secskip\vskip
16pt\parindent=0.8pt\leftskip=18pt plus1fil\parfillskip=0pt
\rightskip=0pt plus1fil\spaceskip=0.3333em\xspaceskip=0.5em
\item{E.}{\bf Spin component measurements}\par\nobreak

\parindent=8.8pt\leftskip=0.8pt\parfillskip=0pt plus1fil\rightskip
=0pt\spaceskip=0em\xspaceskip=0em
\relax In a Stern-Gerlach experiment the direction in which
a spinning particle is deflected depends on the direction of
the force
$$\displayindent=0.8pt\displaywidth=239.7pt F_z=-\nabla\left
(-\mu.B\right)=\mu_z{\partial B_z\,\over\partial z},\eqno(24)$$

where $\mu$ is the magnetic moment of the particle associated
with the spin of the particle and $B$ is the magnetic field
of the magnet, having a strong gradient $\partial B_z/\partial
z$. As seen from the particle{\tt\char13}s point of view, depending
on the force, the pole where the magnetic field is strongest
is either pushed away or attracted by the particle as it passes
through the gap between the poles. This effect is similar to
the force exerted on a geodesic test particle in the model space-time.
If we look at Eq. (22a) we can see that the $1/r$ potential
gives rise to either an attractive force or a repelling force,
depending on the sign of $X$. This close analogy is the reason
why we can say that the sign of $X$ corresponds to the outcome
of a spin component measurement.

\vskip0pt plus\secskip\penalty-250\vskip0pt plus-\secskip\vskip
16pt\parindent=0.4pt\leftskip=18pt plus1fil\parfillskip=0pt
\rightskip=0pt plus1fil\spaceskip=0.3333em\xspaceskip=0.5em
\item{IV.}{\bf CONCLUSION}\par\nobreak

\vskip 10pt\parindent=8.8pt\leftskip=0.4pt\parfillskip=0pt plus1fil\rightskip
=0pt\spaceskip=0em\xspaceskip=0em
\relax If we accept that we can not know for sure whether space-time
is flat at the scale (in space and time) at which the Bohm-Aharonov
experiment is performed, then we can not evade the conclusion
that Bell{\tt\char13}s and similar proofs, which are all based
on counterfactual statements, do not apply. Non-local angles
and other geometric relations over some distance involving counterfactual
set-ups are not measurable and have, even from a realist point
of view, no definite values, because non-local geometric relations
in curved space-time require operational definitions, which
are of course not applicable to counterfactual set-ups.

A proper description of the EPR and Bohm-Aharonov experiments
amalgamates the indefiniteness of the traditional quantum description
with the realist point of view of the traditional classical
description. Such a description is in the spirit of general
relativity and makes plausible that, contrary to common opinion,
the philosophical foundation of general relativity is also fundamental
to the proper interpretation of quantum mechanics as a statistical
theory in a non-flat and largely unknown playground.

An analysis of the results of the Bohm-Aharonov experiment (which
are assumed to agree with the predictions of quantum mechanics)
indicates that a hidden variable can be introduced to explain
the results, but that the configuration of three directions
(the orientations of the measuring instruments and the direction
of the hidden variable) has one degree of freedom less than
expected classically. Classically, given two of the three angles
that identify a configuration, the third angle can be chosen
freely from a continuous spectrum. On the other hand, configurations
that are compatible with the predictions of QM restrict the
third angle to a bivalent choice. This restriction is not severe
and does not introduce non-locality by itself: even classically
the third angle is restricted.

A model based on the simple, even naive assumption that spin
has to do with space time structure with rotational symmetry
in one direction, exhibits geodesic movements with fewer orbit
defining constants than are needed for a Keplerian orbit of
a test particle in a central force field. Of the three constants,
only one constant defines the orientation of the orbital plane.
Relative to an orbital plane, any other orbital plane can, up
to a bivalent choice, be specified with a single number. Thus
the model has exactly the required property to ensure that the
predictions of quantum mechanics can be reproduced. 

\vskip0pt plus\secskip\penalty-250\vskip0pt plus-\secskip\vskip
18pt\fontsetaa\parindent=0pt\leftskip=18pt plus1fil\parfillskip
=0pt\rightskip=0pt plus1fil\spaceskip=0.3333em\xspaceskip=0.5em
\item{}\par\nobreak

\fontsetac\vskip 6pt\leftskip=0pt\parfillskip=0pt plus1fil\rightskip
=0pt\spaceskip=0em\xspaceskip=0em
[1] A. Einstein, B. Podolsky \& N. Rosen, Phys. Rev. {\bf47},
777 (1935)

[2] N. Bohr, Phys. Rev, {\bf48}, 696 (1935)

[3] D. Bohm and Y. Aharonov, Phys. Rev. {\bf108}, 1070 (1957)

[4] J.S. Bell, Physics {\bf1}, 195 (1964)

\bye